\documentclass[trackchanges]{aastex701}

\usepackage{graphicx} 
\usepackage{amsmath}
\usepackage{hyperref}
\usepackage{xspace}
\usepackage{ulem}

\usepackage{tikz}
\usetikzlibrary{arrows.meta, positioning, fit, backgrounds}

\usepackage{acronym}
\acrodef{EM}[EM]{electromagnetic}
\acrodef{CBC}[CBC]{compact binary coalescence}
\acrodef{GW}[GW]{gravitational wave}
\acrodef{H0}[$H_0$]{the Hubble constant}
\acrodef{photoz}[photo-$z$]{photometric redshift}
\acrodef{SFR}[SFR]{star formation rate}
\acrodef{SM}[SM]{stellar mass}
\acrodef{Z}[Z]{metallicity}
\acrodef{SED}[SED]{spectral energy distribution}
\acrodef{KDE}[KDE]{Kernel Density Estimate}
\acrodef{LOS}[LOS]{line-of-sight}
\acrodef{GWTC}[GWTC]{gravitational wave transient catalogue}
\acrodef{MOC}[MOC]{multi-order coverage}
\acrodef{CMB}[CMB]{cosmic microwave background}
\acrodef{BNS}[BNS]{binary neutron star}
\acrodef{BBH}[BBH]{binary black hole}
\acrodef{NSBH}[NSBH]{neutron star-black hole}
\acrodef{LVK}[LVK]{LIGO-Virgo-KAGRA Collaboration}
\acrodef{OIM}[OIM]{observationally-informed map}

\usepackage{xcolor}

\newcommand{\mth}[0]{$m_\text{th}$\xspace}
\newcommand{\gwcosmo}[0]{\texttt{gwcosmo}\xspace}

\newcommand{\Hepsmarg}[0]{$H_0 = {71.9}_{-7.5}^{+9.1}$ km s$^{-1}$ Mpc$^{-1}$}

\begin{document}

\title{Expanding the scope of dark siren cosmology: Inferring the population properties of gravitational wave-hosting galaxies}

\author[orcid=0000-0002-5556-9873]{Rachel Gray}
\affiliation{SUPA, School of Physics and Astronomy, University of Glasgow, Glasgow G12 8QQ, UK}
\email[show]{Rachel.Gray@glasgow.ac.uk}
\author[orcid=0000-0003-3772-198X]{Daniel Williams}
\affiliation{SUPA, School of Physics and Astronomy, University of Glasgow, Glasgow G12 8QQ, UK}
\email[]{Daniel.Williams@glasgow.ac.uk}
\author[orcid=0009-0006-1882-996X]{Alexander Papadopoulos}
\affiliation{SUPA, School of Physics and Astronomy, University of Glasgow, Glasgow G12 8QQ, UK}
\email[]{a.papadopoulos.1@research.gla.ac.uk}

\begin{abstract}
    The field of gravitational wave ``dark siren'' cosmology is facing a number of challenges surrounding the way in which galaxy catalogue data is incorporated into the analysis. The three biggest challenges are adequately compensating for galaxy catalogue incompleteness, robust treatment of galaxy redshift and source properties, and the as-yet unknown host galaxy weighting model. We present an updated methodology for inferring cosmological parameters using dark sirens and incomplete galaxy catalogues. By switching to a sampled redshift prior, rather than one which is pre-computed on a grid, we can jointly infer parameters which describe the population of gravitational-wave hosting galaxies, alongside the gravitational wave population and cosmological model. Using the cosmological inference software \gwcosmo, we reproduce a subset of GWTC-5.0 cosmology results, then extend the analysis to jointly infer the host galaxy weighting model---the first measurement of its kind to do so---leading to an updated measurement of the Hubble constant with \Hepsmarg (median and 68\% symmetric credible interval). This innovative new method opens the door to solving several of the biggest challenges currently facing the field of dark siren cosmology.
\end{abstract}

\section{Introduction \label{sec:intro}}
\Acp{GW} from \acp{CBC} were first proposed as a new cosmological probe in the 1980s~\citep{Schutz:1986}. Their signals encode within them the luminosity distance to the source, a self-calibrated measurement that is entirely independent of the cosmic distance ladder. The modern-day Hubble tension emerged with Planck's 2013 \ac{CMB} measurement \citep{Planck2013}, and has strengthened to a level of $5\sigma$ in the time since \citep{DiValentino_2021}. Whether this is due to unaccounted for systematics in local measurements of \ac{H0}, or a missing ingredient in the cosmological model assumed by early-Universe measurements, is as-yet uncertain. The need for alternative probes of the Hubble tension has reached an all-time high.

During the second observing run of the advanced detector era, the field of \ac{GW} cosmology took off in practice with the 2017 detection of a \ac{BNS} and its \ac{EM} counterpart which allowed for direct identification of the source's host galaxy, and hence its redshift \citep{GW170817:discovery, GW170817:MMA}.
Combined with the luminosity distance of the source this led to the first ever measurement of \ac{H0} from \acp{GW} \citep{GW170817:H0}. Forecasts predicted that 50-100 of these \textit{bright sirens} would produce a measurement precision of 2\%: precise enough to shed light on the ongoing tension \citep{Chen:2017rfc, Feeney:2018mkj}.
However, despite this promising start, the \ac{LVK}'s third and fourth observing runs both passed with no further confirmed \ac{EM} counterparts to any \acp{GW} detected since, despite a significant increase in detector sensitivity. 
Not only have \ac{EM} counterparts been absent, but the sources which can produce them (\acp{BNS}, or low-mass \ac{NSBH} mergers) have simply not been observed at the rates initially predicted \citep{GWTC-3,catalog-gwtc-4d0,catalog-gwtc-5}.

Although the rate of bright sirens remains low, the observed rate of \textit{dark sirens} (\acp{GW} observed without an \ac{EM} counterpart, of which the majority are \acp{BBH}) have exceeded initial predictions and are arriving in sufficient quantity that they may hold the key to reaching a 2\% measurement of \ac{H0} within the next decade. 
For dark sirens, there are two places where the redshift information necessary for a cosmological measurement enters the analysis:
\begin{enumerate}
    \item The component masses of compact objects inferred from \ac{GW} signals are increased by a factor of $(1+z)$ from the source frame, as a result of cosmological redshift, $z$. 
    While relatively uninformative at the individual-detection level, at population-level this effect plays an important role due to features in the \ac{CBC} mass distribution which break the mass-redshift degeneracy \citep{1993ApJ...411L...5C,PhysRevD.48.4738,Taylor:2011fs,Farr:2019twy, PhysRevD.104.062009,Agarwal_2025}.
    With increased observation time by the \ac{LVK} detectors, the number of \ac{GW} detections will grow; the precision possible through this method alone will thus continue to increase. 
    Where this is the sole source of redshift information entering the analysis, the term \textit{spectral sirens} (referring to the mass distribution, or ``\textit{spectrum}'') is commonly used \citep{PhysRevLett.129.061102}.
    \item In addition to redshift information from the \ac{GW} population, galaxy catalogues can be used to provide the redshift of the \textit{potential} host galaxies within the \ac{GW}'s inferred localisation volume. 
    By statistically marginalising over the uncertainty of which galaxy is the true host, the contributions from the true host galaxies of multiple \acp{GW} constructively combine to provide a tighter constraint on \ac{H0} \citep{Schutz:1986,MacLeod:2007jd,DelPozzo:2012,Chen:2017rfc,Fishbach:2018gjp,Gray:2019ksv,Gair_2023}.\footnote{We note that there is an alternative method of utilising gravitational waves and galaxy catalogues for cosmology is that of cross-correlation, which due to differences in methodology is not subject to the same set of systematics as dark siren cosmology \citep[see, e.g.][]{PhysRevD.93.083511,2021PhRvD.103d3520M,2020ApJ...902...79B,Mukherjee_2024,peaksirens}, though we do not consider this method further in this paper}
\end{enumerate}

The amount of information a single dark siren contributes cosmologically depends strongly on two things: how well-localised the \ac{GW} is; and how complete the galaxy catalogue is, and the quality of the redshifts it contains, within that localisation volume. 
\citet{Borghi_2024} forecasts a 1\% measurement of \ac{H0} with the 100 loudest \acp{GW} detections from one year of the \ac{LVK}'s fifth observing run (O5), assuming a \textit{complete} galaxy catalogue with spectroscopic-like redshift uncertainties.\footnote{Here ``complete'' means ``containing all possible potential host galaxies'', rather than the more common meaning in \ac{EM} communities of ``complete [to some apparent magnitude]''.}
While optimistic, this study highlights both the importance that small localisation volumes have to play (see also investigations into the constraining power of ``gold'' and ``silver'' dark sirens, e.g. \citet{Borhanian_2020,Dang2026}), and the importance of precise redshift measurements (the same study with photometric-like redshift uncertainties returns a measurement of \ac{H0} degraded by a factor of $\sim 9$).

In order to make a future percent-level measurement of \ac{H0} with dark siren a reality, there are three major challenges on the \ac{EM} side:
\begin{enumerate}
    \item \textit{Galaxy catalogue incompleteness.} Galaxy catalogues do not have uniform coverage over the sky for many reasons. 
    They are also not complete (in a volume-limited sense) over the redshift range in which we detect \acp{CBC}. 
    There are two options for accounting for this. 
    The first is to apply cuts to the catalogue to reduce it to a volume-limited sample which can be utilised with a subset of \acp{GW}~\citep[see e.g.][]{Soares-Santos:2019irc,GW190814:DES,Palmese_2023}. 
    The second is to apply an incompleteness correction to account for the missing galaxies~\citep[see e.g.][]{Gray:2019ksv,Gray2022,Gray_2023, Mastrogiovanni2023, Finke:2021aom,Borghi_2024,Tagliazucchi2025}.\\
    \hspace*{1em}Dark siren methods which apply a galaxy catalogue incompleteness correction typically do so with a pixelated approach, which allows for a directionally-dependent completeness correction to be applied. 
    The redshift prior is then pre-computed per-pixel, which allows for costly incompleteness calculations to be done in advance of the hierarchical cosmological inference. 
    While computationally effective and memory efficient, these pre-computed \ac{LOS} redshift priors contain hardcoded assumptions about the host galaxy weighting model and the incompleteness correction that has been applied.\\
    \hspace*{1em}Incompleteness corrections to date typically work by either modelling the \ac{EM} selection cut of the galaxy catalogue using an apparent magnitude threshold, and then adding on an incompleteness correction that is computed assuming astrophysical priors on redshift and luminosity; or estimating the number of observed galaxies within a certain volume, compared to the expected number for a complete catalogue, and ``filling in'' the missing galaxies. 
    Both methods require assumptions about the astrophysical distribution of galaxies which could hypothetically contribute to the dark siren analysis---not just the observed ones.
    
    \item \textit{Robust treatment of galaxy redshift and source properties.} Galaxy redshift catalogues do not provide the full posteriors for every galaxy they contain, as doing so would require an infeasible amount of storage. 
    Instead this information is compressed into statistics, for example the mean and standard deviation, or confidence intervals. 
    If the values of other galaxy source properties are also reported, such as luminosity or stellar mass, which is not guaranteed, those too are compressed, and information about   correlations between these parameters and redshift are lost. 
    Analyses using the full galaxy posteriors were applied in \citep{GW190814:DES, Palmese_2023,Bom_2024} using a subset of \ac{GW} detections and a volume limited catalogue, but so far methods which additionally compensate for (redshift dependent) galaxy catalogue incompleteness have not done so, losing potentially important correlations between parameters.
    
    \item \textit{An unknown host galaxy weighting model.} The astrophysics which govern compact binary formation and evolution tells us that the host galaxy environment plays an important role in the types of binaries that form and the rates at which they do so \citep[see e.g.][]{Mapelli_2018,Neijssel_2019, Adhikari_2020, Santoliquido_2021, Broekgaarden_2022, Srinivasan_2023, Rauf_2023}). 
    While this depends on the host galaxy's properties, such as \ac{SM}, \ac{SFR}, and metallicity, the relationship is poorly defined and has large uncertainties associated with it. 
    Dark siren analyses to date have typically utilised a range of luminosity-based weightings (in various bands, which can act as tracers for the properties mentioned above), but for each analysis this host weighting model has been fixed. There have been several studies investigating the impact of assuming the incorrect host galaxy weighting model, where the scale of the impact depends strongly on how far the chosen tracer deviates from the underlying truth \citep{Hanselman_2025, Perna_2025, Alfradique_2025}.
\end{enumerate}

In summary, there has yet to be a dark siren method presented that can compensate for galaxy catalogue incompleteness, utilise full galaxy posteriors on redshift and (potentially correlated) source properties, and can marginalise over the unknown host galaxy weighting model. In this paper, we present a methodology that enables all three of these things.
By switching to a redshift prior which is represented by a set of \textit{samples}, rather than one which is pre-computed on a grid, we gain the flexibility to address each of the challenges listed above.
 
In Section \ref{sec:method}, we present the derivation of the updated method, followed by details of the implementation within the cosmological inference software \gwcosmo~\citep{Gray:2019ksv,Gray2022,Gray_2023,Papadopoulos2026}.\footnote{Available at \href{https://git.ligo.org/lscsoft/gwcosmo}{https://git.ligo.org/lscsoft/gwcosmo}} In section~\ref{sec:GWTC-5_analysis} we present the analysis of the fifth Gravitational Wave Transient Catalogue (GWTC-5.0) with the GLADE+ galaxy catalogue using the new method both to demonstrate consistency with results from~\citep{gwtc5:cosmo}, and carry out a new analysis which additionally infers parameters of the host galaxy weighting model. In Section~\ref{sec:discussion} we discuss the implications of the new method for the field of dark siren cosmology, in light of the three challenges outlined above. And finally, in Section~\ref{sec:conclusions}, we conclude.

\section{Method \label{sec:method}}

\subsection{Bayesian framework}
Below we outline a hierarchical Bayesian framework which which infers a set of cosmological and population hyperparameters, $\Lambda$, using a set of $N_\text{det}$ \ac{GW} observations and a redshift prior which is constructed from a galaxy catalogue.
The set of hyperparameters, $\Lambda$ is composed of three subsets: $\Lambda_\text{cosmo}$, the parameters of the model's cosmology; $\Lambda_{\text{pop}}$, the population parameters which define the underlying astrophysical distribution of compact binaries in the model, and their distribution in redshift, $z$; and $\Lambda_{\text{gal}}$, the parameters of the astrophysical distribution of galaxies which contain \acp{CBC} in the model.

\acp{CBC} are characterised by a number of both intrinsic and extrinsic astrophysical parameters which can be inferred from the observed gravitational-wave strain data, $x_\text{GW}$ which are a timeseries composed of noise and (potentially) and additive signal.
The noisy nature of these data, or equivalently, the finite sensitivity of the \ac{GW} detector, result in some \ac{GW} signals being undetectable; the parameter $D_{\text{GW}}$ is a binary parameter in our model which indicates whether the \ac{GW} data contain a trigger which passed some detection threshold statistic in order to be deemed \emph{detected}. 

The posterior on $\Lambda$ can be constructed as follows \citep{Mandel:2018mve,Vitale:2020aaz}:
\begin{equation}\label{eq:hierarchical}
\begin{aligned}
p(\Lambda|\{x_\text{GW}\},\{D_\text{GW}\},I)\ &\propto \ p(\Lambda|I)\ p(N_\text{det}|\Lambda,I)\  p(D_{\text{GW}}|\Lambda,I)^{-N_\mathrm{det}}\ \prod^{N_\text{det}}_i p(x_{\text{GW}i}|\Lambda,I),
\end{aligned}
\end{equation}
with $\{x_\text{GW}\}$ the set of strain data for all analysed \acp{GW} in the model, and $\{D_{\text{GW}}\}$ the set of their corresponding detection statuses.




For a dark siren analysis incorporating galaxy catalogue data, we can expand this expression to explicitly show where the redshift information will enter. 
Galaxy surveys are inherently flux-limited, so we must account for their (in-)completeness. 
The correction we must apply for this completeness is not uniform across the sky, due to survey boundaries and the inability to survey galaxies behind the Milky Way.
\citet{Gray2022} developed an approach which segmented the sky into a set of pixels, $\{\Omega_j\}$, to address this, where the prior probability of any \ac{GW} originating in the part of the sky covered by a given pixel is $p(\Omega_j\ |\ \Lambda, I)$.
This allows us to expand the term $p(x_{\text{GW}i}\ |\ \Lambda, I)$ as

\begin{equation}\label{Eq:sumpix}
\begin{aligned}
p(x_{\text{GW}i}\ |\ \Lambda, I) = \sum^{N_\mathrm{pix}}_{j=1} p(x_{\text{GW}i}\ |\ \Omega_j, \Lambda, I) \ p(\Omega_j\ |\ \Lambda, I).
\end{aligned}
\end{equation}

If we assume an isotropic and homogenous universe, $p(\Omega_j\ |\ \Lambda, I)$ is equivalent to the fractional area of the sky covered by the pixel. Previous works
\citet{Gray2022,Gray_2023}~assumed equally-sized pixels covering the sky, so this term was constant; here we retain this term explicitly for the additional flexibility of allowing a scheme with pixels of different sizes, which we will refer to as the \emph{multi-resolution scheme}.

\begin{table}[]
    \centering
    \begin{tabular}{|c|l|}
        \hline
       \textbf{Symbol}  &  \textbf{Definition} \\
       \hline
       $x_\text{GW}$  & GW strain data \\
       $D_\text{GW}$  & Binary flag: is the \ac{GW} signal in this section of data detected? \\
       $\Lambda$  & $\{\Lambda_\text{cosmo}, \Lambda_\text{pop}, \Lambda_\mathrm{gal}\}$, the set of cosmological and population hyperparameters \\
       $\vec{\phi}$  &  Galaxy source parameters \\
       $z$ & Cosmological redshift \\
       $G$ ($\bar{G}$)  & Binary flag: is this galaxy contained within the galaxy catalogue? \\
       \hline
    \end{tabular}
    \caption{A list of parameters appearing in the dark siren methodology and their definitions}
    \label{tab:parameterdef}
\end{table}

The likelihood of having observed the data, $x_{\text{GW}i}$, given the origin pixel $\Omega_j$, a given redshift, $z$, and the set of model hyperparameters, $\Lambda$, is $p(x_{\text{GW}i}|z, \Omega_j, \Lambda, I)$.
The joint prior on the redshift and galaxy parameters is $p(z, \vec{\phi}|\Omega_j,\Lambda,I)$.
We ultimately wish to obtain the likelihood in terms only of the origin pixel and the model hyperparameters, so we can marginalise over $\vec{\phi}$, and $z$:

\begin{equation}\label{Eq:sum_likelihood_compact}
\begin{aligned}
p(x_{\text{GW}i}|\Omega_j, \Lambda, I) &= \iint p(x_{\text{GW}i}|z, \Omega_j, \Lambda,I) \ p(z,\vec{\phi}|\Omega_j, \Lambda,I) \; \text{d}z \, \text{d}\vec{\phi}.
\end{aligned}
\end{equation}



The prior on $z$ is necessarily joint with the prior on the galaxy properties, $\vec{\phi}$.
The properties, $\vec{\phi}$, of a galaxy at a given $z$ will determine the probability that a \ac{CBC} would originate from it, and whether the galaxy will be included in the catalogue.
In \cite{Gray_2023}, $\vec{\phi} = M$, where $M$ is the galaxy absolute magnitude (in a given observing band). 
In reality, $\vec{\phi}$ could be any number of galaxy source parameters such as \ac{SM}, \ac{SFR} and \ac{Z}.  
We use the term $\vec{\phi}$ for generality.  
Following a similar derivation to \cite{Gray_2023}, from here on we explicitly include the term $s$ on the RHS of the probability distributions to indicate that these are conditioned on a \ac{GW} source being present, as well as GC to indicate the galaxy catalogue data.
The probability that a galaxy with properties $\vec{\phi}$ at a redshift $z$ will be the source of a \ac{CBC} is $p(s|\, \vec{\phi}, I)$: the host galaxy weighting model.
The joint prior on $z$ and $\vec{\phi}$ is then
\begin{equation}
\begin{aligned}
p(z,\vec{\phi}\ |\ \Omega_j,\Lambda,\mathrm{GC},s,I) \propto p(s\ |\ z,\vec{\phi},I) \ p(z,\vec{\phi}\ |\ \Omega_j,\Lambda,\mathrm{GC},I).
\end{aligned}
\end{equation}
We can ignore the normalisation of this equation by recognising that the prior formulation we outline here will be utilised in the denominator to compute \ac{GW} selection effects, and so constants which can be taken outside of the integrals will later cancel. 
The exact form of $p(z,\vec{\phi}|\Omega_j,\Lambda,\mathrm{GC},I)$ depends on the type of galaxy catalogue in use.
 
If the galaxy catalogue is \emph{complete} (meaning it is a volume-limited catalogue that contains all possible host galaxies for the detected set of \acp{GW}), $p(z,\vec{\phi}|\Lambda,\mathrm{GC},I)$ can be constructed directly from the galaxy catalogue with no need to correct for any incompleteness.
For a catalogue containing $N_\text{G}(\Omega_j)$ galaxies in pixel $\Omega_j$, and in the absence of measurement uncertainties, this is simply
\begin{equation}
\begin{aligned}\label{Eq:sumgal_nouncert}
p(z,\vec{\phi}\ |\ \Omega_j,\Lambda,\mathrm{GC},I) = \dfrac{1}{N_\text{G}(\Omega_j)} \sum^{N_{\text{G}}(\Omega_j)}_{k=1} \delta(z - z_{k}) \delta(\vec{\phi} - \vec{\phi}_{k}),
\end{aligned}
\end{equation}
that is, a sum of delta functions corresponding to each galaxy within the pixel.
Generally, if $p(z,\vec{\phi}\ |\ \Omega_j,\Lambda,\mathrm{GC},I)$ can be expressed purely as a sum of weighted delta functions, then using the Monte Carlo integration technique with this prior allows equation~\ref{Eq:sum_likelihood_compact} to be evaluated as
\begin{equation}
\begin{aligned}
p(x_{\text{GW}}\ |\ \Omega_j,\Lambda,s,I) &\propto \dfrac{1}{N_{\text{G}}(\Omega_j)} \sum^{N_{\text{G}}(\Omega_j)}_{k=1} p(x_{\text{GW}}\ |\ z_{k},\Omega_j,\Lambda,I)\  p(s\ |\ z_{k},\vec{\phi}_{k},I).
\end{aligned}
\end{equation}

Equation~\ref{Eq:sumgal_nouncert} is an idealised scenario: in reality there will be some measurement uncertainty on both $z$ and $\vec{\phi}$ (though it may be close to negligible for $z$ if the redshift has been determined from spectroscopy).  
Assuming that $z$ and $\vec{\phi}$ have been inferred using a Bayesian analysis, and that $N_{\text{s}, k}$ samples are generated from the posterior probability distribution generated by that analysis, corresponding to the $k$-th galaxy, we can use these  to construct the prior:
\begin{equation}
\begin{aligned}\label{Eq:sumgal}
p(z,\vec{\phi}|\Omega_j,\Lambda,\mathrm{GC},I) = \dfrac{1}{N_G(\Omega_j)} \sum^{N_G(\Omega_j)}_{k=1} \left[ \dfrac{1}{N_{\text{s},k}} \sum^{N_{\text{s},k}}_{l=1} \delta(z - z_{k,l}) \delta(\vec{\phi} - \vec{\phi}_{k,l}) \right].
\end{aligned}
\end{equation}
We have assumed here that consistent assumptions were used when inferring $z$ and $\vec{\phi}$ to those which will be used in the rest of the dark siren analysis.
If not, the posterior samples can generally be reweighted to remove their original prior and apply a consistent one.

If the galaxy catalogue is \textbf{not} complete, an extra stage must be included in the formulation of the redshift prior.
We must marginalise over the probability that a given galaxy is, $p(G| z,\vec{\phi},\Omega_j,\mathrm{GC},I)$, or is not, $p(\bar{G}| z,\vec{\phi},\Omega_j,\mathrm{GC},I)$, contained within the galaxy catalogue.
We introduce the new notation for the prior constructed in equation~\ref{Eq:sumgal}, $p(z,\vec{\phi}|G, \Omega_j, \Lambda,\mathrm{GC},I)$, with this term having the same form as previously, but now representing only the case that the galaxy is within the catalogue.
The complementary prior, $p(z,\vec{\phi}|\bar{G}, \Omega_j, \Lambda,\mathrm{GC},I)$, represents the opposite case, where the galaxy is not contained within the catalogue.

The prior which combines these two cases is
\begin{equation}
\begin{aligned}\label{Eq:inoutcat}
p(z,\vec{\phi}\ |\ \Omega_j, \Lambda,\mathrm{GC},I) = p(z,\vec{\phi}\ |\ G,\Omega_j, \Lambda,\mathrm{GC},I)\ p(G\ |\ \Omega_j, \Lambda, \mathrm{GC}, I) + p(z,\vec{\phi}\ |\ \bar{G}, \Omega_j, \Lambda,\mathrm{GC},I)\ p(\bar{G}\ |\ \Omega_j,\Lambda, \mathrm{GC}, I).
\end{aligned}
\end{equation}

To compute $p(z,\vec{\phi}|\bar{G}, \Omega_j, \Lambda,\mathrm{GC},I)$ we apply Bayes theorem:
\begin{equation}
\begin{aligned}
p(z,\vec{\phi}|\bar{G}, \Omega_j, \Lambda,\mathrm{GC},I) = \dfrac{p(\bar{G}| z,\vec{\phi}, \Omega_j, \mathrm{GC},I) p(z,\vec{\phi}|\Lambda,I)}{p(\bar{G}|\Omega_j,\Lambda,\mathrm{GC},I)}.
\end{aligned}
\end{equation}
The denominator will cancel with the equivalent expression in equation~\ref{Eq:inoutcat}, leaving only the two terms in the numerator to compute explicitly. 
Computing $p(\bar{G}| z,\vec{\phi},\Omega_j,\mathrm{GC},I)$ requires us to compute selection statistics of the galaxy, $\vec{\rho}$, from the provided source parameters, $\vec{\phi}$. 
For example, in~\citet{Gray_2023}, $\rho=m(M,z,\Lambda)$, the apparent magnitude which is computed from the galaxy absolute magnitude, redshift, and a fixed set of cosmological parameters. 
The galaxy catalogue threshold is defined on $\rho$, and is used to down-select the samples that would \textit{not} be observed by the galaxy survey, and hence not be included in the catalogue. 
We can represent this term with a Heaviside step function, $\Theta(\rho(z,\vec{\phi}) -\rho_\text{th}(\Omega_j))$. 
More generally, $\vec{\rho}$ are the statistics used to define whether or not a galaxy is included in the final catalogue, such as apparent magnitude in any number of bands, or the galaxy colour. 
Importantly, $\vec{\rho}_\text{th}$, may be a function of $\Omega_j$, allowing for different thresholds to be applied in different patches of the sky.

The second term $p(z,\vec{\phi}|\Lambda,I)$ is the prior probability distribution on $z$ and $\vec{\phi}$, with no information from the galaxy catalogue.
In \citet{Gray_2023} this was taken to be uniform in co-moving distribution over $z$, and a Schechter function for $\vec{\phi}$ \citep{Schechter:1976}, but in actuality this could be any (sensible) prior on both $z$ and whatever parameters have been selected for $\vec{\phi}$.
In order reduce the computational demands of evaluating the integrals over $z$ and $\vec{\phi}$, samples may be drawn directly from $p(z,\vec{\phi}|\Lambda, I)$ and used in a Monte Carlo integration.

The $p(G|\Omega_j,\Lambda, \mathrm{GC}, I)$ term, is simply the proportion of galaxies within a given pixel which are contained within the catalogue to all (observed or non-observed) galaxies which are within that pixel:
\begin{equation}
\begin{aligned}\label{Eq:probincatreal}
p(G|\Omega_j,\Lambda, \mathrm{GC}, I) \equiv \frac{N_G(\Omega_j)}{N_\mathrm{all}(\Omega_j)},
\end{aligned}
\end{equation}

$N_\mathrm{all}(\Omega_j)$ is not accessible, however, by assuming a number density of galaxies, $n^*(z, \vec{\phi})$ its approximate value can be determined as
\begin{equation}
\begin{aligned}\label{Eq:Nall}
N_\mathrm{all}(\Omega_j) = p(\Omega_j|I) \iint n^*(z,\vec{\phi})\ p(z,\vec{\phi}|\Lambda, I)\ \text{d}z\, \text{d}\vec{\phi},
\end{aligned}
\end{equation}
While this approach requires the assumption of a galaxy density, when constructing the final \ac{LOS} redshift prior, the term $N_G(\Omega_j)$ cancels for the in-catalogue term. 
This makes the approach robust in the limit of arbitrarily-high pixel resolutions without requiring $N_G(\Omega_j)$ to be computed at a coarser resolution (as was the case in \cite{Gray_2023}), thus ensuring the correct relative weighting of galaxies in different pixels.

The full expression for the redshift prior in a single pixel is then
\begin{equation}
\begin{aligned}\label{Eq:inoutexpanded}
p(z,\vec{\phi}|\Omega_j,\Lambda,\mathrm{GC},I) &= \dfrac{1}{N_\mathrm{all}(\Omega_j)}  \sum^{N_\text{G}(\Omega_j)}_{k=1} \left[ \dfrac{1}{N_{\mathrm{s},k}}\sum^{  N_{\mathrm{s},k}}_{l=1} \delta(z - z_{kl}) \delta(\vec{\phi} - \vec{\phi}_{k,l})\right]  \\  &\quad +  \Theta(\rho(z,\vec{\phi}) - \rho_\mathrm{th}(\Omega_j)) \dfrac{1}{N_\mathrm{samp}}\sum^{N_\mathrm{samp}}_{m=1} \delta(z - z_{m}) \delta(\vec{\phi} - \vec{\phi}_{m}).
\end{aligned}
\end{equation}

In practise we construct this prior as a list of $N$ samples with corresponding weights.
The weight, $w_n$, of the $n$-th sample is thus given by:
\begin{equation} \label{eq:weights}
w_n(\Lambda) = \begin{cases}
\dfrac{1}{N_\mathrm{all}(\Omega_j) N_{\mathrm{s},k}} , & \text{if sample $n$ was drawn from the galaxy catalogue,}\\
 \dfrac{1}{N_\mathrm{samp}} \times \Theta(\rho_\mathrm{th}(\Omega_j) - \rho(z_m,\vec{\phi}_m)) , & \text{if  sample $n$ was drawn from the completeness correction.}
\end{cases}
\end{equation}
Again, this assumes that the initial priors that the samples were drawn with are the ones which should be used for cosmological inference. 
If, however, the priors need to be changed (for self consistency, or, indeed, if the hyperparameters of those priors are being jointly inferred as part of the inference), then the weights must additionally include the ratio of the new and old prior evaluated at the sample's location.

The gravitational wave likelihood, $p(x_{\text{GW}}|\Omega_j,\Lambda,s,I)$, can therefore be expressed as
\begin{equation}\label{eq:GWlikelihood}
\begin{aligned}
p(x_{\text{GW}}|\Omega_j,\Lambda,s,I) &\propto  \sum^{N}_{n=1}  p(x_{\text{GW}}|z_{n},\Omega_j,\Lambda,I) p(s|z_{n},\vec{\phi}_{n},I) w_n.
\end{aligned}
\end{equation}

\subsection{Implementation within \gwcosmo}

The method outlined above introduces a huge deal of flexibility to dark siren cosmology (and \ac{GW} population analyses in general), but comes with one major downside relative to the method as presented in \cite{Gray_2023}. 
In that paper, where the \ac{LOS} prior is stored in a grid (referred to as the \textit{gridded \ac{LOS} prior} from here on), the overall \textit{size} of the redshift prior is proportional to the number of grid points used. 
Judicious summing of pixels means that this grid resolution (and the total number of \ac{GW} events being analysed) is the dominant deciding factor for the amount of memory required, which is entirely \textit{independent} of the size of the original galaxy catalogue.

In this new method (which we will refer to as the \textit{sampled \ac{LOS} prior}), because samples are stored for each galaxy, a naive implementation leads to the memory scaling directly with the number of galaxies in the galaxy catalogue (as well as the number of \acp{GW} used). 
In order to combat this, we introduce the use of a multi-resolution pixelation scheme.
We define the pixel sizes based on the coverage of \ac{GW} events, such that smaller pixels are used to cover the most well localised regions, and poorly localised areas (or areas with no \ac{GW} coverage at all) are covered by larger pixels. 
Where the pixel area is greater than some user-defined threshold, samples are drawn purely from the agnostic $p(z,\vec{\phi}|\Lambda, I)$ prior, under the understanding that structure in pixels with volumes much larger than the scale of homogeneity will average out, meaning that the gain from incorporating real galaxy catalogue data in these parts of the sky is minimal. 
This significantly reduces the total number of pixels required, and limits the total number of samples per pixel so it no longer scales directly with pixel area. Thus the benefits of incorporating the full \ac{GW} population into a cosmological analysis are retained, at the fraction of the cost.

Below we summarise the steps to generate the \ac{LOS} prior in more detail, which are specific to the implementation, though we note that these are likely to be refined in the future. 

\subsubsection{Creating an observationally-informed map}\label{sec:oim}

To make the analysis computationally tractable, we must generate a multi-resolution map (also known as a \ac{MOC} map) which identifies the optimal resolution scheme to use in order to capture every well-localised \ac{GW} which falls within the galaxy catalogue footprint. 
We utilise a binary flag for pixels where $1=\mathrm{GOOD}$ (the \ac{LOS} prior for this pixel will be computed using the galaxy catalogue), and $0=\mathrm{BAD}$ (the \ac{LOS} prior for this pixel will be computed without information from the galaxy catalogue, i.e. only containing the astrophysical prior). Because this map is constructed based on both \ac{GW} and galaxy observations, we call it the \ac{OIM}.

The \ac{OIM} generation is done in three stages:
\begin{enumerate}
    \item \textbf{Generation of a \ac{GW} multi-resolution map.} The purpose of this map is to identify patches of the sky which are covered by well-localised \acp{GW}, and the optimal pixelation grid to represent them. 
    The inputs used to generate it are the \ac{GW} skymaps of every \ac{GW} that is planned to be used in the analysis. 
    We implement a user-defined threshold on three quantities. 
    The first is the maximum amount of \ac{GW} probability which can be contained in any given pixel. 
    This criterion determines the resolution with which each \ac{GW} is covered. 
    Starting at the highest possible resolution, pixels are summed until they meet this stopping criterion. 
    This leads to well-localised \acp{GW} being covered with higher-resolution pixels. \\
    \hspace*{1em}The other two quantities then define which of these pixels are labelled as GOOD.  
    One is the maximum pixel size (\texttt{nside}), under the assumption that galaxy structure in pixels larger than this will be averaged-out, and therefore would make a minimal contribution to the final posterior anyway. 
    The other is the minimum probability for any one pixel to contain. 
    If a \ac{GW} is poorly localised, such that none of its pixels pass the both thresholds, that event is tagged with `treat as spectral', such that when later passed to \gwcosmo, that event will be analysed with the astrophysical (non-galaxy informed) prior. 
    \item \textbf{Generation of a galaxy catalogue multi-resolution map.} The purpose of this map is to identify patches of the sky with galaxy catalogue coverage, and within this, utilise a pixelation scheme which groups patches of the sky with similar levels of completeness. 
    The input for this is a precomputed map of galaxy number density, or a precomputed map of the galaxy apparent magnitude threshold (\mth; it should be noted that a magnitude threshold map is still a necessary input to gwcosmo when computing the full \ac{LOS} prior---see Section \ref{sec:LOSprior_implementation}). 
    To facilitate computation of the ideal pixelation grid, we implement two user-defined criteria. 
    The first is the minimum (maximum) value on galaxy number (\mth) that a pixel should have to be labelled as GOOD, at the resolution of the input map. 
    The second is maximum relative difference that the values in a set of sub-pixels can have, to be merged to a coarser resolution, (i.e. a threshold on $(\mathrm{max}-\mathrm{min})/\mathrm{mean}$). 
    The aim here is to both identify patches of the sky where the galaxy catalogue has coverage, and identify sharp boundaries where the completeness of the galaxy catalogue changes.
    \item \textbf{The union of the GW and galaxy catalogue multi-resolution maps.} The final stage of constructing the \ac{OIM} is to add the GW and galaxy multi-resolution maps together, thereby creating a pixelation regime that captures both the well-localised \acp{GW} and the galaxy survey boundaries.
    Pixels which were labelled GOOD in both input maps retain the label of GOOD, and every other pixel is labelled BAD. 
    The pixelated grid can then be simplified by combining BAD pixels, as these will not be utilised. Pixels which contain 0 are merged to successively lower resolutions until they hit a resolution that would include a GOOD pixel.
\end{enumerate}
The \ac{OIM} is then utilised for the generation of the sampled \ac{LOS} prior, which is detailed in Section \ref{sec:LOSprior_implementation}.

\subsubsection{Generating and storing the LOS redshift prior~\label{sec:LOSprior_implementation}}

The creation of the \ac{LOS} prior requires as inputs the \ac{OIM}, the galaxy catalogue data (currently limited to, for every galaxy, the redshift mean, standard deviation, right ascension, declination, and apparent magnitude in a used-specified band), a pixelated map which quantifies the completeness of the catalogue across the sky (defined in terms of a directionally dependent \mth; particularly important now that the multi-resolution approach enables pushing to very high resolutions, such that any method of computing galaxy catalogue completeness based on the contents of a specific pixel may be unreliable), and the chosen astrophysical prior $p(z,\vec{\phi}|\Lambda, I)$ (where for now $\vec{\phi} = M$, absolute magnitude). We anticipate a number of significant upgrades to this in the near future to enable a more flexible approach to ingesting galaxy catalogue data.

We use the \ac{OIM} described in section~\ref{sec:oim} to dictate the grid on which the prior is to be computed; this allows the overall size of the prior to be minimised, as minimal data is stored for uninformative areas of the sky. 

We define and then store the \ac{EM} selection criteria for each pixel in the \ac{OIM} by referencing the provided \mth map; the grid of this map is independent of the pixelation defined by the \ac{OIM}. Higher resolution pixels in the \ac{OIM} inherit the \mth of the corresponding parent pixel in the magnitude threshold map.
An additional criteria based on redshift allows the user to define cuts in redshift which can be used to discard galaxy redshift samples which are deemed to be unreliable.

In each pixel of the \ac{OIM} we determine the prior independently of the pixel's resolution; for GOOD pixels this is built from a mixture of two terms, as described in equation~\ref{Eq:inoutexpanded}: the in-catalogue and the completeness correction terms. 
We draw samples from each of these independently, and store these separately. 
This allows the samples to be used later for per-event weighting during inference.
Redshift samples for the in-catalogue term are drawn from a normal distribution with a mean and standard deviation determined by the galaxy's recorded redshift and uncertainty from the catalogue, truncated to exclude $z<0$.
An absolute magnitude value is also computed for each sample, derived from the recorded apparent magnitude of the galaxy in the catalogue, the sample's redshift, and any necessary $K$-correction and evolution correction. Samples which do not pass the pixel's selection criteria are discarded.
For the completeness-correction term, samples are drawn from the astrophysical priors on redshift and galaxy source properties. No cuts based on the \ac{EM} selection criteria are applied to the completeness-correction samples at this stage, such that the completeness-correction samples in every pixel are consistent with the astrophysical prior $p(z,\vec{\phi}|\Lambda, I)$, which provides a pool of samples which can be used to evaluate the \ac{GW} likelihood in the BAD pixels, or for \acp{GW} which have been tagged with ``treat as spectral''. For BAD pixels, no galaxy data is utilised and we only store the completeness-correction samples.

When called by \gwcosmo, the \ac{LOS} prior for a specific pixel is constructed instantaneously by returning the in-catalogue and completeness-correction samples and their corresponding weights computed as in Eq. \ref{eq:weights}.

In order to evaluate \ac{GW} selection effects, we additionally require a sky-averaged \ac{LOS} prior which can be evaluated with a Monte Carlo integral with a set of search sensitivity injections. While \gwcosmo currently only has separate (non correlated) priors on $z$ and $M$ implemented, we can pre-compute this quantity and store it as an interpolatable array. We emphasise that once a correlated astrophysical prior is implemented, $p(z,\vec{\phi}|\Lambda, I)$, this will need to be upgraded so that the sky-averaged prior can be reweighted with the same ease as the \ac{LOS} prior.



\subsection{The dark siren likelihood}

As the prior is sample-based, and the \ac{GW} data utilised is also sample-based (posterior samples, to be specific), the \ac{GW} data needs to be smoothed in order for the Monte Carlo integration shown in Eq. \ref{eq:GWlikelihood} to be carried out. We use the same approach as in \citet{Gray2022,Gray_2023} of constructing a set of \acp{KDE} over the \ac{GW} redshift distribution, each corresponding to a different patch of the sky. These \acp{KDE} are regenerated for each likelihood evaluation (incorporating the marginalisation over other \ac{GW} properties which change with each new draw of $\Lambda$).

While the method remains broadly the same, the implementation has been improved. We compute a single \ac{KDE} for every pixel in a \ac{GW} footprint which has been labelled as GOOD using all of the \ac{GW} samples which fall in that pixel. As the \ac{OIM} has been generated to based on some maximum probability $p$ per-pixel, this limits the total number of pixels required to cover any \ac{GW} to of order $1/p$ (the actual number will be higher than this, as this is strictly a minimum). The majority of pixels contain a similar amount of \ac{GW} probability, and therefore a similar number of posterior samples. Thus the choice of max \ac{GW} probability per pixel correlates closely to the number of posterior samples that will be used to generate the \acp{KDE}.
For samples which lie in pixels which are labelled BAD, these are now grouped together and a single \ac{KDE} is utilised to represent this distribution, which is then evaluated using purely the astrophysical prior on $z$ and $\vec{\phi}$. The prior samples are taken from a random pixel's completeness-correction samples, and are different for each \ac{GW} to avoid exacerbating any random structure. For \acp{GW} which have been tagged with ``treat as spectral'', only single \ac{KDE} is generated, and evaluated using the astrophysical prior in a similar manner.
Each \ac{KDE} is weighted by the total \ac{GW} sky probability contained within the pixel(s) it corresponds to.

\section{Analysis of GWTC-5.0 with the GLADE+ galaxy catalogue \label{sec:GWTC-5_analysis}}

\subsection{Data}
We utilise data from GWTC-5.0~\citep{catalog-gwtc-1,catalog-gwtc-2,catalog-gwtc-2d1,GWTC-3,catalog-gwtc-4d0,catalog-gwtc-5}, using the same selection threshold as the corresponding cosmology paper~\citep{cosmo-gwtc-5d0} (false alarm rate $<\,0.25\,\mathrm{yr}^{-1}$), yielding a set of 235 dark sirens in our analysis. 
This event set does not include the bright siren GW170817, which is treated separately. 
We also use the publicly-available sensitivity injection campaigns for quantifying GW selection effects.
For all analyses, we use the \textsc{FullPop-4.0} population model \citep{Fishbach_2020,Farah_2022,Mali_2025} and parametrise the redshift evolution of the GW events using a Madau-Dickinson distribution \citep{Madau_2014}. 
When marginalising over the GW population parameters we employ prior ranges consistent with those found in~\citet{cosmo-gwtc-5d0}.

For the galaxy catalogue, we utilise $K$ band galaxies from the GLADE+ galaxy catalogue \citep{GLADE+}, which is a composite catalogue  made from cross-matching multiple surveys. When converting galaxy apparent magnitudes to absolute magnitudes we apply $K$-corrections following the prescription in \citet{Kochanek_2001}. To model the incompleteness correction we assume a $K$-band Schechter function with $M^* = -23.39 + 5 \log(h)$, where $h \equiv H_0/100$, $\alpha = -1.09$, a faint-end Schechter cut at $M_\mathrm{max} = -19.0 +5 \log(h)$, and $\phi^* = 1.16 \times10^{-2} \ h^3 \mathrm{\ Mpc}^{-3}$ \citep{Kochanek_2001}. The GLADE+ $K$-band contains about 1.16 million sources, and shows a significant drop off in completeness by $z=0.1$, which means that it is not informative for the bulk of the GWTC-5.0 detections which lie at distances beyond this. However, its use in \citet{cosmo-gwtc-5d0} and \citet{Gray_2023} makes it useful for comparison, and a number of \ac{GW} detections have occurred within its footprint at low enough redshifts for it to significantly enhance the information they contribute to the dark siren analysis.

\subsection{The LOS redshift prior}
\begin{figure}
    \centering
    \includegraphics[width=0.9\linewidth]{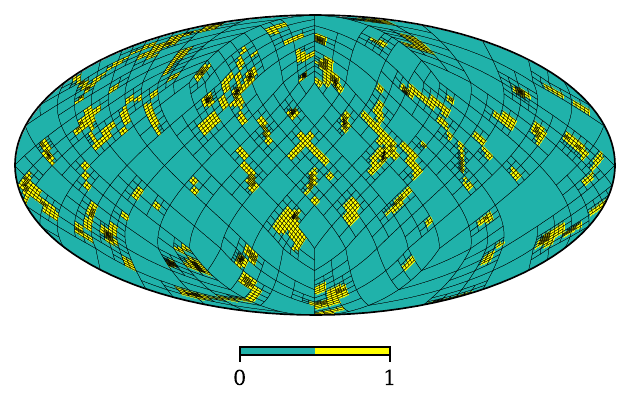}
    \caption{The MOC mask for GWTC-5.0 and the GLADE+ $K$-band catalogue, with a maximum \ac{GW} probability per-pixel limit of 5\%, and a minimum nside of 32 for GOOD (yellow) pixels. Gridlines denoting pixel boundaries are shown in black.}
    \label{fig:moc_mask}
\end{figure}

We generate the sampled LOS redshift prior according to the method outlined in Section \ref{sec:method}. 
We utilise a maximum \ac{GW} probability per-pixel limit of 5\%, and a minimum nside of 32 to determine the resolution of the \ac{OIM}, and determine which pixels meet the GOOD criteria. The final \ac{OIM}, shown in Figure \ref{fig:moc_mask}, contains 3330 pixels, of which 2265 are labelled GOOD, and the highest nside it contains is 256 (this highest resolution is dictated by the choice of a 5\% max \ac{GW} probability per pixel - decreasing this percentage would force higher resolutions, at the expense of reducing the number of \ac{GW} posterior samples used to set up the \ac{GW} \ac{KDE} in each pixel).

\begin{figure}
    \centering
    \includegraphics[width=\linewidth]{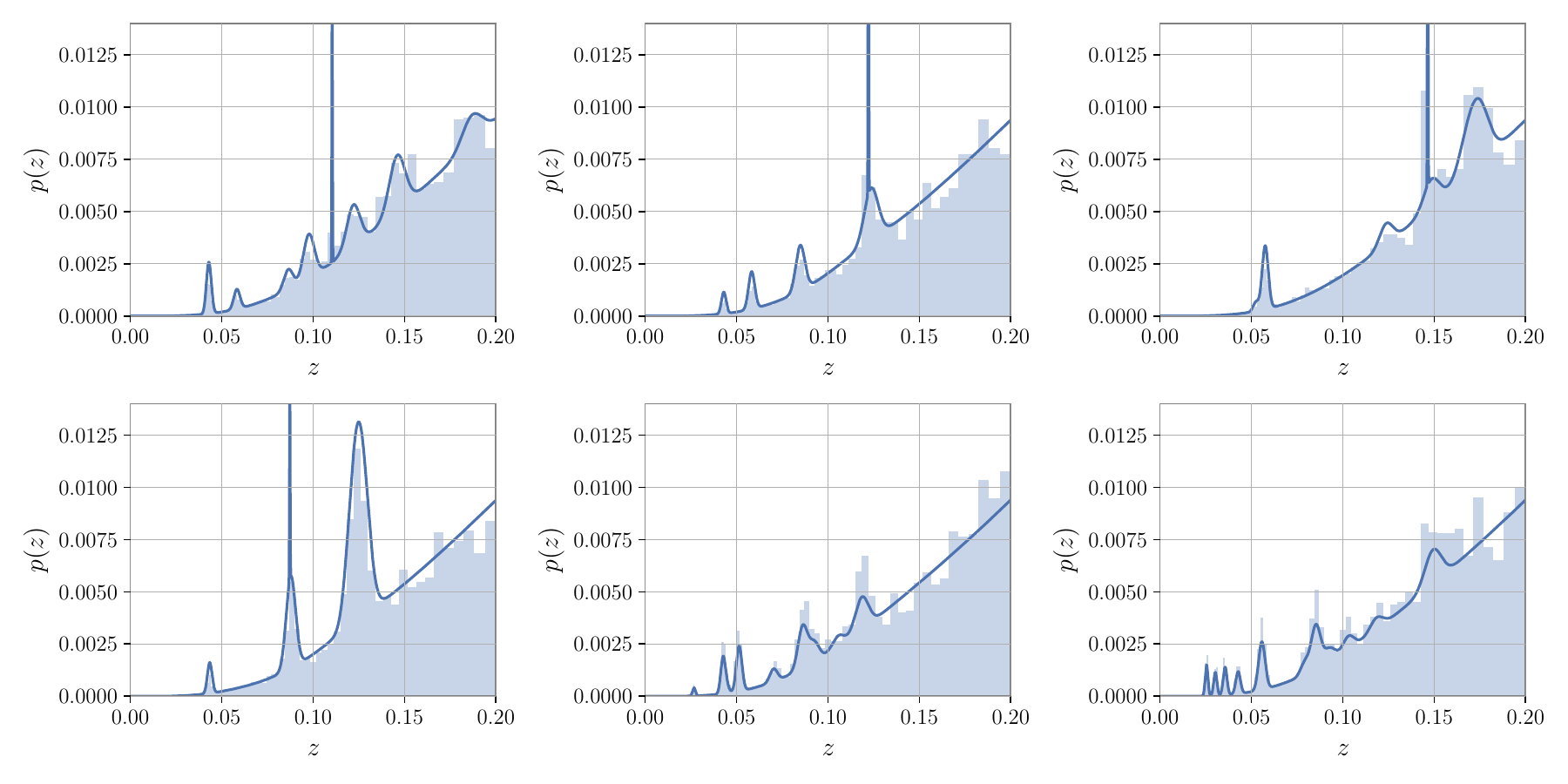}
    \caption{Comparison of six nside 64 GOOD pixels generated using the gridded LOS prior method (solid line) and the sampled LOS prior method (histogram). Host galaxy weighting is proportional to galaxy luminosity.}
    \label{fig:LOSprior_pixels_comparison}
\end{figure}

We check for consistency with \ac{LOS} priors generated using the gridded approach. 
Because the sampled \ac{LOS} prior is multi-resolution, we choose to compare a subset of pixels which are of the same order as a pre-existing gridded \ac{LOS} prior, corresponding to nside 64. 
Figure~\ref{fig:LOSprior_pixels_comparison} shows six pixels which meet the nside=64 criteria and are also labelled GOOD in the sampled prior.
The agreement is broadly good, with galaxy structure captured, but in different pixels the sample-based prior can have more or less support than the gridded case, due to the switch from using the observed number of galaxies in each pixel to normalise the in-catalogue contribution, to using the theoretical $N_\mathrm{all}$. 
Nside 64 pixels typically have fewer than 50 galaxies for the GLADE+ $K$-band, leading to noticeable up- or down-weighting of the in-catalogue contribution due to small number statistics.

\begin{figure}
    \centering
    \includegraphics[width=0.7\linewidth]{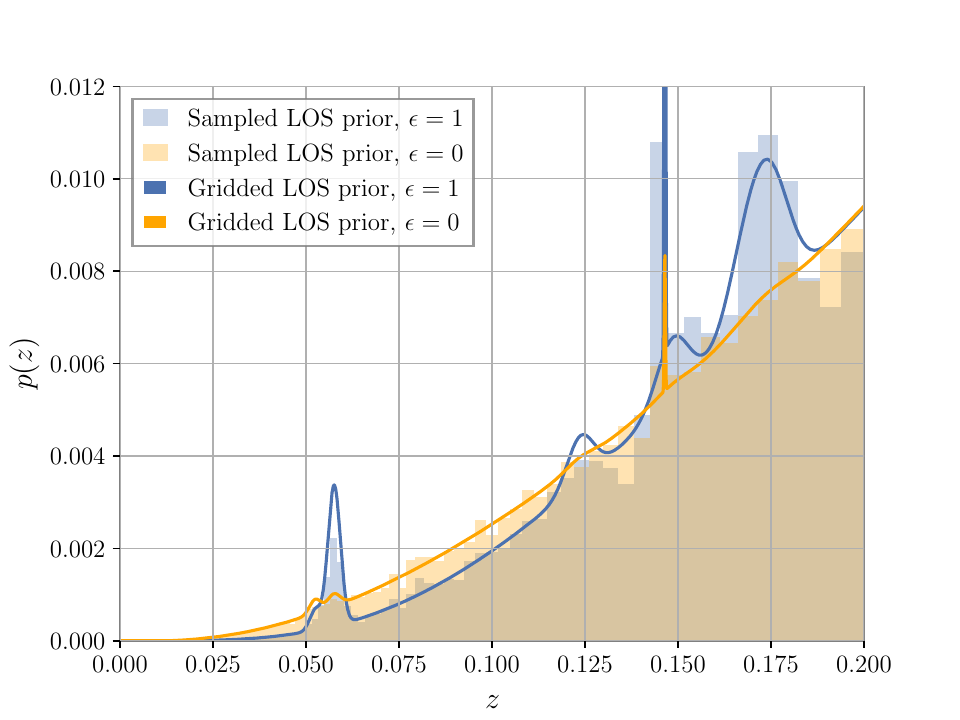}
    \caption{Comparison of an nside 64 GOOD pixel using the gridded LOS prior method (solid line) and the sampled LOS prior method (histogram) for a host weighting model with $\epsilon=0$ (uniform weights; orange) and $\epsilon=1$ (weights proportional to luminosity; blue).}
    \label{fig:LOSprior_weights_comparison}
\end{figure}

Figure~\ref{fig:LOSprior_weights_comparison} shows a single pixel assuming different host weighting model choices: $\epsilon=1$ (weighted proportional to luminosity) and $\epsilon=0$ (uniform weighting). The same set of samples is utilised in both weighting cases, and simply reweighted by the host galaxy weighting model, $\propto L^\epsilon$. Again, reasonable agreement is demonstrated with the gridded \ac{LOS} prior.

\subsection{Results}
To demonstrate the efficacy of the new sampled \ac{LOS} redshift prior, we present a set of analyses using single \ac{GW} events while fixing \ac{GW} population parameters, as well as the full GWTC-5.0 catalogue while jointly inferring population parameters. Where possible we compare results with those computed using the gridded \ac{LOS} prior (generated with a fixed resolution corresponding to nside 128).

\begin{figure}
    \centering
    \includegraphics[width=\linewidth]{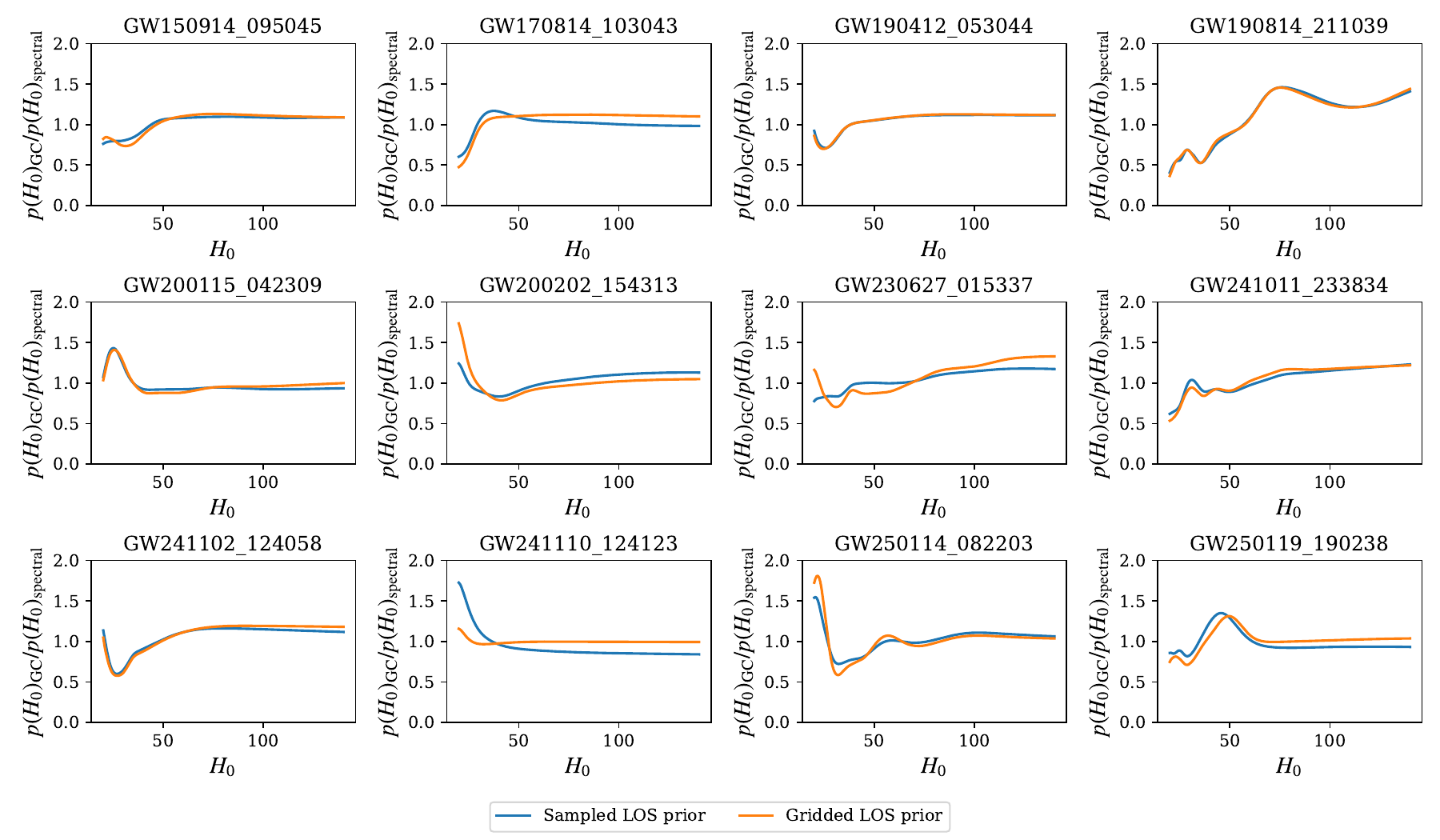}
    \caption{Single event posteriors computed with the sampled (blue) and gridded (orange) \ac{LOS} priors, divided through by the spectral posteriors, assuming fixed population parameters.}
    \label{fig:single_event_ratios}
\end{figure}

For the fixed-population runs, we analysed all \acp{GW} which were not tagged with ``treat as spectral'', assuming a luminosity weighted host galaxy model. The 12 \acp{GW} which show the most significant deviation from the spectral result are shown in Fig. \ref{fig:single_event_ratios}, where we have taken the ratio of the posterior using the catalogue-informed prior, to the spectral posterior. Thus, the deviation of the line from 1 denotes the contribution of the galaxy catalogue. Overall we see that similar structure is picked out using both the gridded and sampled \ac{LOS} priors. There are some deviations, which reflect the differences seen in the \ac{LOS} priors themselves (see e.g. Fig. \ref{fig:LOSprior_pixels_comparison}). The change to multi-resolution pixels for the sampled \ac{LOS} prior also means that the \acp{KDE} which are constructed to represent the 3-dimensional \ac{GW} localisation volume will be different between the two analyses, and could lead to minor changes to the single event posteriors.

\begin{figure}
    \centering
    \includegraphics[width=0.5\linewidth]{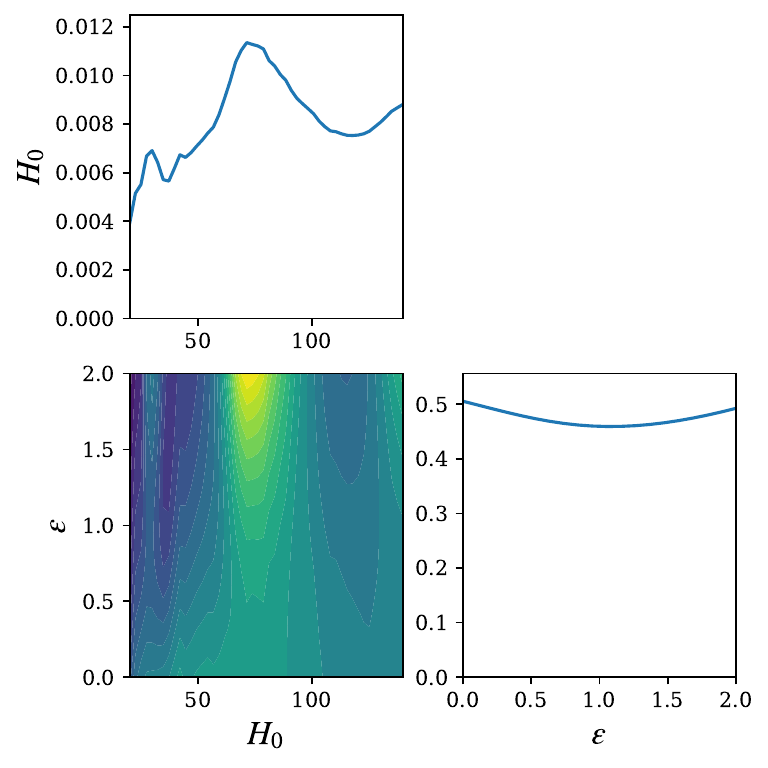}
    \caption{Corner plot for GW190814 showing $H_0$ and $\epsilon$, with all other population parameters fixed.}
    \label{fig:GW190814_grid}
\end{figure}

The dark siren GW190814\_211039 shows the most significant deviation from the spectral result across the full range of the prior, so we carry out a 2-dimensional analysis for this event, jointly inferring $H_0$ and $\epsilon$. The results are shown in Fig. \ref{fig:GW190814_grid}. While relatively uninformative on $\epsilon$, this analysis does confirm that the amount of structure in the $H_0$ posterior is correlated to $\epsilon$, with a low $\epsilon$ corresponding to a relatively uniform distribution on $H_0$, and larger values exacerbating the structure caused by the galaxy catalogue.

\begin{figure}
    \centering
    \includegraphics[width=\linewidth]{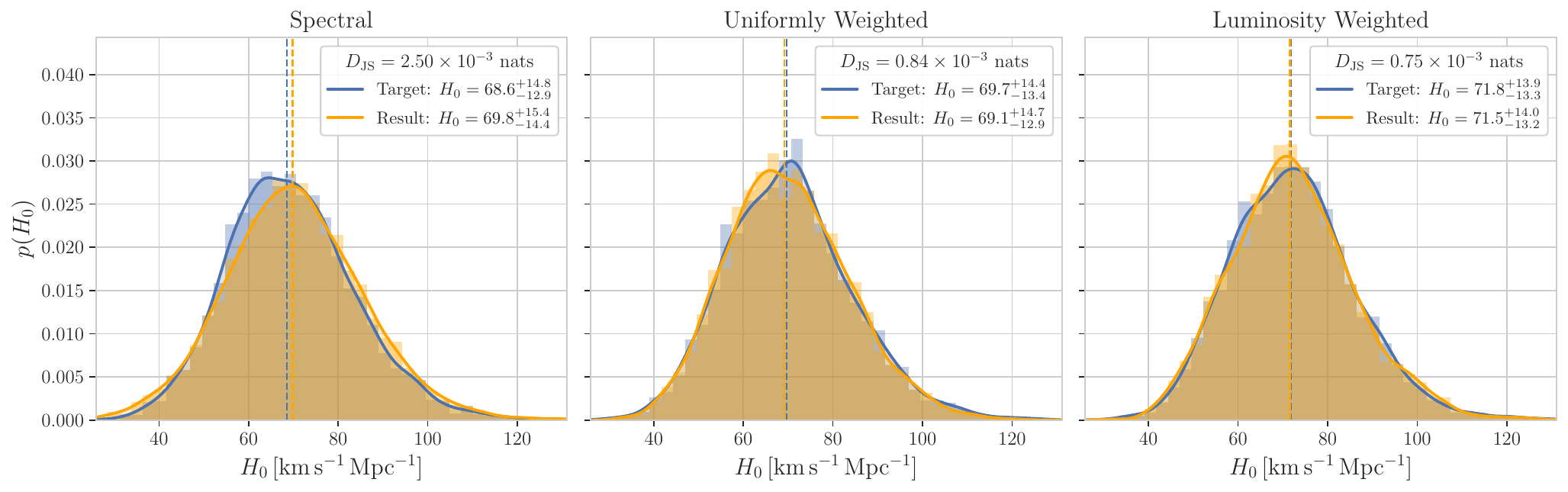}
    \caption{A comparison of the $H_0$ posteriors from  the GWTC-5.0 analyses using the gridded \ac{LOS} prior (blue) and the sampled \ac{LOS} prior (orange), marginalised over population parameters.}
    \label{fig:GWTC-5_comparison}
\end{figure}

Having confirmed that the single-event posteriors demonstrate the expected behaviour, we then look at results for the full GWTC-5.0 dataset. We compare the results from spectral, uniform-weighted ($\epsilon=0$) and luminosity-weighted ($\epsilon=1$) using both the sample-based and the gridded LOS prior. The results are shown in Fig. \ref{fig:GWTC-5_comparison}. It shows the $H_0$ posteriors produced by different analyses when marginalising over the CBC population parameters. Considering dark siren analyses with uniform and luminosity weighting as well as spectral siren only analyses, this demonstrates good agreement found between the new sample-based LOS approach against the previous pre-computed method. We also compute the Jensen-Shannon (JS) divergence \citep{jsd} between the sampled and gridded results for each scenario. Considering the agreement criterion from \citet{RomeroShaw2020}, we find that there is statistical agreement between the uniformly weighted and luminosity weighted case, while the spectral case just passes the 0.002 nat threshold to imply a significant deviation. Given the differences in implementation which could give rise to changes in the posterior, discussed further in section \ref{sec:conclusions}, we do not believe this deviation to be truly significant, or indicative of an error in the sampled \ac{LOS} prior method.

\begin{figure}
    \centering
    \includegraphics[width=\linewidth]{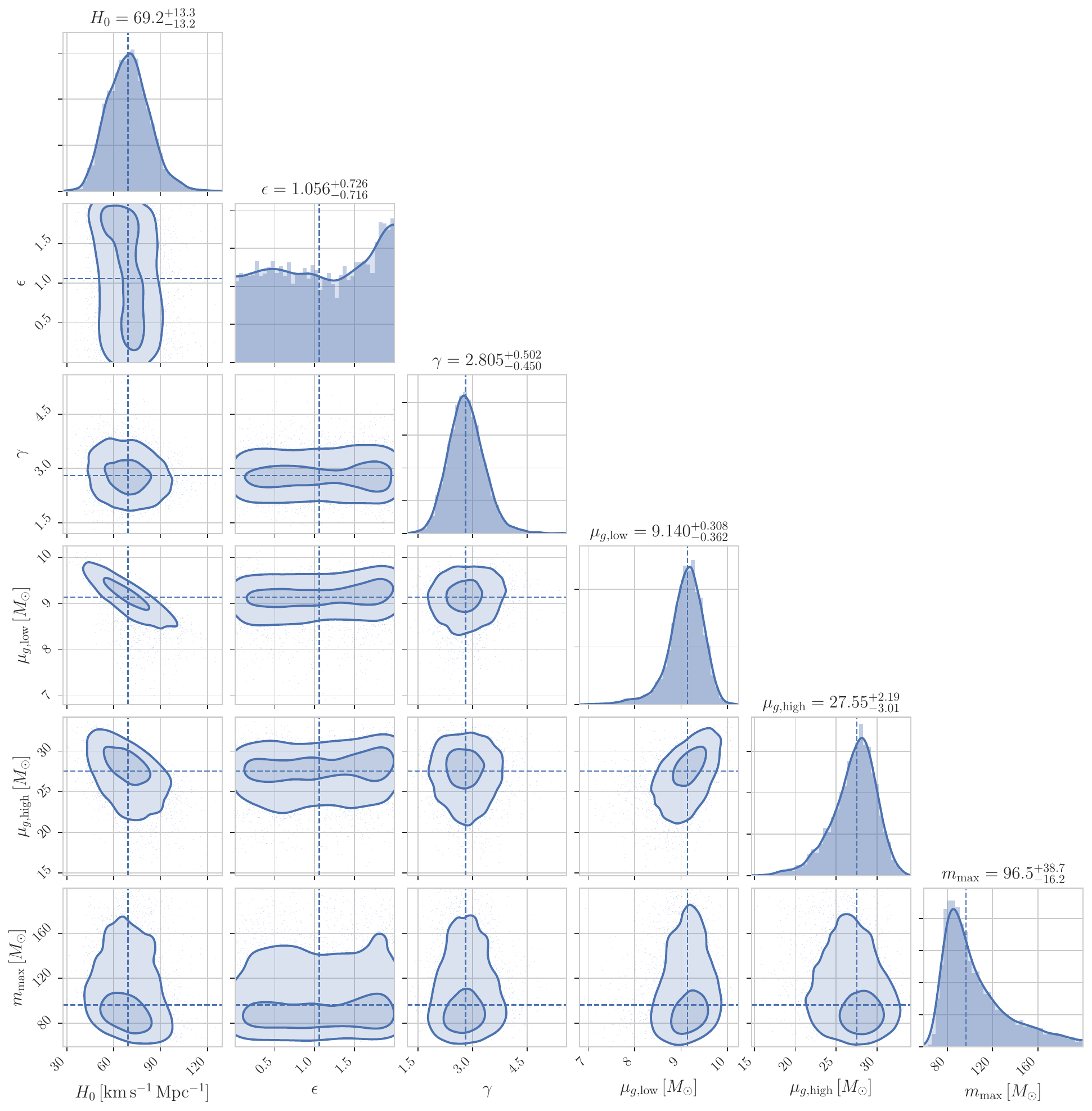}
    \caption{A corner plot showing a subset of parameters for the GWTC-5.0 analysis with GLADE+, assuming a flexible host galaxy weighting model parameterised by $\epsilon$. Contours show the $1\sigma$ and $2\sigma$ credible intervals.}
    \label{fig:GWTC-5_corner}
\end{figure}

Finally, we carry out an analysis marginalising over $\epsilon$, assuming a uniform prior in the range $\left[0,2\right]$. This is the first analysis of its kind which jointly infers parameters of the host galaxy weighting model alongside cosmological and \ac{GW} population parameters. Fig. \ref{fig:GWTC-5_corner} is a reduced corner plot showing the correlations between $H_0$, $\epsilon$, and the population parameters which correlate most strongly with $H_0$. $\epsilon$ does not show any particular correlation with any other parameters. It is poorly constrained over the prior range, which is unsurprising due to the very low levels of completeness the GLADE+ galaxy catalogue has in the $K$-band. Combining the marginal posterior on $H_0$ with the constraint from the bright siren GW170817 leads to an updated value of \Hepsmarg.

\section{Discussion \label{sec:discussion}}

\subsection{Assessment of current results}
The lack of completeness in the GLADE+ $K$-band limits the precision reachable using GWTC-5.0. The host-galaxy weighting parameter, $\epsilon$, in particular, is poorly constrained, and results which marginalise over it should be treated with caution due to the strong prior-dependence that they retain. While not correlated with $H_0$ in such a way to systematically shift the peak of the posterior, larger values of $\epsilon$ \textit{do} correlate with increased structure in the \ac{LOS} prior, which result in a more tightly constrained measurement of $H_0$. Combined with the fact that increasing $\epsilon$ reduces the number of effective samples in the \ac{LOS} prior, we caution against pushing $\epsilon$ to high values as, dependent on the dataset, this could artificially narrow the $H_0$ posterior. With a deeper galaxy catalogue, and larger numbers of \acp{GW} with small localisation volumes, however, it should be possible to constrain $\epsilon$ and remove this risk.

While the results using the sampled \ac{LOS} prior are generally in agreement with those from the gridded \ac{LOS} prior, it is notable that the spectral result, in which \textit{no} information comes from the galaxy catalogue, are not perfectly in agreement. There are a couple of reasons that this could be. The first is that, with the gridded \ac{LOS} prior, the spectral analysis is still carried out in a pixelated fashion, and involves summing multiple pixels to marginalise over the sky. With the sampled \ac{LOS} prior, however, the spectral analysis effectively treats the whole sky as one pixel, and so the \ac{GW} data never gets split and recombined.\footnote{It should be noted that this is not inherently due to the gridded vs sampled nature of the prior - rather the upgrades that were necessary to implement the sampled prior naturally led to an improvement to how the spectral case was handled in \gwcosmo.} This is not anticipated to cause a systematic shift to the $H_0$ posterior, but it should be noted that a detailed systematics study on the effect of this splitting and recombining has not yet been carried out, and remains a task for the future. We anticipate that the new method \textit{should} be more robust to any effects that this could cause, as \ac{GW} posterior samples which fall outside GOOD pixels are grouped together, rather than being unnecessarily split and recombined.

The other plausible change that could cause a systematic shift to the posterior at the population level is how the redshift prior is applied when computing \ac{GW} selection effects. The choice to model the prior as a simple uniform in co-moving distribution, rather than averaging the \ac{LOS} prior over the whole sky, as is done in the gridded case, could have an impact. Because the evaluation of the integral over redshift is taken to the power of the number of detected events (see Eq. \ref{eq:hierarchical}), minor deviations may stack up in a noticeable fashion. We do not expect this to have a significant impact here due to the relatively low completeness of the galaxy catalogue (meaning that a deviation from uniform in co-moving would only happen at very low redshifts, corresponding to a relatively small fraction of the overall volume being integrated over), but we caution that this is something that should be investigated in detail in the future.

\subsection{Implications for the future}

We have demonstrated the ability of this method to recreate previous results and marginalise over a simple host galaxy weighting model. However, the true aim with this method is to move beyond the simplistic assumptions that have been used in dark siren cosmology to date. Referring back to the three main challenges highlighted in Section \ref{sec:intro}:
\begin{enumerate}
    \item \textit{Galaxy catalogue incompleteness.} The sample-based \ac{LOS} prior enables us to move away from the case where the \ac{EM} selection criteria is modelled as a single apparent magnitude threshold, and instead can handle selection effects that depend on multiple observation bands and galaxy colour (for example). To make this a reality, it must be possible to move from a set of galaxy properties (redshift, and whichever source properties have been selected, $\vec{\phi}$) to a predicted set of \textit{observables} through a forward model.
    \item \textit{Robust treatment of galaxy redshift and source properties.} When utilising photometric redshifts, galaxy posterior samples on redshift and source properties can be used directly, avoiding potential sources of bias that come from neglecting the correlation of these parameters, or assuming them to have Gaussian uncertainty. When utilising spectroscopic redshifts, infinitely high redshift precision is possible without vastly increasing memory requirements.
    \item \textit{An unknown host galaxy weighting model.} The sampled \ac{LOS} prior can be reweighted on the fly to apply any host galaxy weighting model so long as it is consistent with the galaxy source properties chosen, meaning that this host galaxy weighting model can be jointly inferred alongside cosmological and \ac{GW} population parameters. Additionally, the ability to flexibly update the astrophysical priors used for the ``out of catalogue'' completeness correction means that effectively we are able to infer the \textit{population} of \ac{CBC}-hosting galaxies.
\end{enumerate}

While this new method comes with a great deal of promise, its robustness depends strongly on having sufficient samples to accurately represent the \ac{LOS} redshift prior. 
Extreme changes to the host galaxy weighting model or galaxy astrophysical priors will significantly change the number of effective samples the prior contains and could lead to stability issues if not carefully monitored. 
We anticipate that a significant amount of development will have to be carried out in this direction to ensure the stability and robustness of this method as the number of parameters and complexity of the models increases.

Looking to the future, and at further opportunities, this new method may additionally enable estimation of cosmological parameters beyond the Hubble constant, such as parameters describing the dark energy equation of state, which will be of interest to those utilising the dark siren method to put constraints on modified gravity models that effect \ac{GW} propagation \citep[see, e.g.][]{Finke:2021aom, AChen2023}.
However, we caution that while the Bayesian methodology presented here allows for the cosmological model to be reweighted during inference, care will have to be taken to ensure that implicit assumptions about the cosmological model which may be present in the galaxy data and the inferred galaxy redshifts and source properties are correctly accounted for.

Outside the field of cosmology, this method enables further constraints to be applied directly to the population properties of the galaxies that host \acp{CBC} (and these constraints will be significantly more informative if the cosmological model is fixed, rather than jointly inferred). \citet{Vijaykumar_2024} demonstrated that it is possible to use \acp{GW} to distinguish between galaxy host weighting models if the evolution with redshift is sufficiently different between those considered. There, the constraining power comes from modelling the merger rate as a function of redshift, and noting that different weighting models (tracing e.g. \ac{SFR} or \ac{SM}) produce different expected distributions in redshift, rather than utilising data from individual galaxies. In \citet{Li_2026} it was demonstrated that the inclusion of galaxy catalogue data lends additional constraining power, again driven by the most well-localised \acp{GW}.

\section{Conclusions \label{sec:conclusions}}
We have presented an upgrade to the dark siren methodology, which enables joint inference of the \ac{CBC} host galaxy population as well as the \ac{GW} population and cosmology. 
We have implemented the first iteration of this methodology within the cosmological inference software \gwcosmo and demonstrated its ability to recreate results from \cite{gwtc5:cosmo}. 
Additionally, we have presented a new analysis on the GWTC-5.0 dataset, using the GLADE+ galaxy catalogue $K$-band data, this time jointly inferring the parameter which dictates the probability of a galaxy hosting a \ac{CBC}, alongside cosmological and \ac{GW} population parameters, resulting in a measurement of \Hepsmarg. 
This results presented in this paper are intended as a ``proof of principle'' for the new method, but the truly exciting aspect of this work is that it opens the door to solving the three key major challenges that face dark siren cosmology from the \ac{EM} side: galaxy catalogue incompleteness, robust treatment of galaxy redshifts and source properties, and the unknown host galaxy weighting model. 
A huge amount of developmental work remains necessary, which will require input from experts across the fields of \ac{GW} cosmology and populations, \ac{EM} galaxy surveys, and astrophysical modelling. 
With a clear path forward, the goal of reaching a 2\% measurement of \ac{H0} by the mid-2030s is tantalisingly close.

\section*{Acknowledgements}
RG and DW are supported by the Science and Technology Facilities Council (STFC) grant UKRI2487. AP is supported by the UKRI STFC studentship 323353-01. This material is
based upon work supported by NSF’s LIGO Laboratory
which is a major facility fully funded by the National
Science Foundation.

\bibliography{bib}{}
\bibliographystyle{aasjournalv7}

\end{document}